\documentclass[
aps,
prd,
twocolumn,
preprintnumbers,
superscriptaddress,
floatfix,
preprintnumbers,
longbibliography,
nofootinbib
]{revtex4-2}
\usepackage[dvipdfmx]{graphicx}
\usepackage{dcolumn}
\usepackage{bm}
\usepackage{ulem}
\usepackage{amsmath}
\usepackage{amssymb}
\usepackage{txfonts}
\usepackage{hyperref}
\usepackage{color} 
\usepackage{xcolor}
\usepackage{listings}
\usepackage{cprotect}

\usepackage{tikz}
\usetikzlibrary{quantikz2}

\graphicspath{{./figures/}}

\DeclareMathOperator{\tr}{tr}

\newcommand{\e}{\mathrm{e}}

\newcommand{\im}{\mathrm{i}}

\hypersetup{
  colorlinks=true,
  linkcolor=[rgb]{0.60,0.00,0.00},
  citecolor=[rgb]{0.00,0.00,0.60},
  urlcolor=[rgb]{0.00,0.00,0.60},
  setpagesize=false
}

\definecolor{codegreen}{rgb}{0,0.6,0}
\definecolor{codegray}{rgb}{0.5,0.5,0.5}
\definecolor{codepurple}{rgb}{0.58,0,0.82}
\definecolor{backcolour}{rgb}{0.95,0.95,0.92}

\lstdefinestyle{mystyle}{
    backgroundcolor=\color{backcolour},   
    commentstyle=\color{codegreen},
    keywordstyle=\color{magenta},
    numberstyle=\tiny\color{codegray},
    stringstyle=\color{codepurple},
    basicstyle=\ttfamily\footnotesize,
    breakatwhitespace=false,         
    breaklines=true,                 
    captionpos=b,                    
    keepspaces=true,                 
    numbers=left,                    
    numbersep=5pt,                  
    showspaces=false,                
    showstringspaces=false,
    showtabs=false,                  
    tabsize=2
}

\begin{document}

\preprint{RIKEN-iTHEMS-Report-26, YITP-26-02}

\title{
  Onset of thermalization of q-deformed SU(2) Yang--Mills theory on a trapped-ion quantum computer
}

\author{Tomoya Hayata}
\email{hayata@keio.jp}
\affiliation{Departments of Physics, Keio University School of Medicine, 4-1-1 Hiyoshi, Kanagawa 223-8521, Japan}
\affiliation{RIKEN Center for Interdisciplinary Theoretical and Mathematical Sciences (iTHEMS), RIKEN, Wako 351-0198, Japan}
\affiliation{International Center for Elementary Particle Physics and The University of Tokyo, 7-3-1 Hongo, Bunkyo-ku, Tokyo 113-0033, Japan}

\author{Yoshimasa Hidaka}
\email{yoshimasa.hidaka@yukawa.kyoto-u.ac.jp}
\affiliation{Yukawa Institute for Theoretical Physics, Kyoto University, Kyoto 606-8502, Japan}
\affiliation{RIKEN Center for Interdisciplinary Theoretical and Mathematical Sciences (iTHEMS), RIKEN, Wako 351-0198, Japan}

\author{Yuta Kikuchi}
\email{yuta.kikuchi@quantinuum.com}
\affiliation{Quantinuum K.K., Otemachi Financial City Grand Cube 3F, 1-9-2 Otemachi, Chiyoda-ku, Tokyo, Japan}
\affiliation{RIKEN Center for Interdisciplinary Theoretical and Mathematical Sciences (iTHEMS), RIKEN, Wako 351-0198, Japan}

\begin{abstract}

Nonequilibrium dynamics of quantum many-body systems is one of the main targets of quantum simulations. This focus -- together with rapid advances in quantum-computing hardware -- has driven increasing applications in high-energy physics, particularly in lattice gauge theories. However, most existing experimental demonstrations remain restricted to (1+1)-dimensional and/or abelian gauge theories, such as the Schwinger model and the toric code. 
It is essential to develop quantum simulations of nonabelian gauge theories in higher dimensions, addressing realistic problems in high-energy physics. To fill the gap, we demonstrate a quantum simulation of thermalization dynamics in a (2+1)-dimensional $q$-deformed $\mathrm{SU}(2)_3$ Yang-Mills theory using a trapped-ion quantum computer. By restricting the irreducible representations of the gauge fields to the integer-spin sector of $\mathrm{SU}(2)_3$, we obtain a simplified yet nontrivial model described by Fibonacci anyons, which preserves the essential nonabelian fusion structure of the gauge fields. We successfully simulate the real-time dynamics of this model using quantum circuits that explicitly implement $F$-moves. In our demonstrations, the quantum circuits execute up to 47 sequential $F$-moves. We identify idling errors as the dominant error source, which can be effectively mitigated using dynamical decoupling combined with a parallelized implementation of $F$-moves.

\end{abstract}

\date{\today}

\maketitle

\section{Introduction}
\label{sec:introduction}

Understanding the nonperturbative dynamics of quantum field theories is one of the central challenges in modern physics, and has led to remarkable progress in the physics of phase transitions in quantum many-body systems. A seminal example in condensed matter physics is the microscopic theory of superconductivity developed by Bardeen, Cooper, and Schrieffer~\cite{PhysRev.108.1175}, illustrating the power of quantum field theories and the importance of nonperturbative analysis. Nonperturbative effects of quantum field theories are also important in high-energy physics: the fundamental forces between elementary particles are described by a quantum field theory called a nonabelian gauge theory (this is the so-called Standard Model of particle physics~\cite{Schwartz:2014sze}). It is believed that the early universe underwent phase transitions, and accurate computation of the nonperturbative effects of a nonabelian gauge theory is necessary to understand the phase transitions and evolution of our universe.

Lattice gauge theory, pioneered by Wilson~\cite{PhysRevD.10.2445}, provides the most powerful and established first-principles formulation of gauge theories~\cite{DeGrand}. By defining quantum field theories on a discretized spacetime lattice with a finite number of degrees of freedom, it introduces a natural ultraviolet cutoff to regularize the divergence of quantum field theories. Combined with advanced Monte Carlo methods, the lattice formulation has been remarkably successful in computing a wide range of nonperturbative phenomena in quantum chromodynamics (QCD), the nonabelian gauge theory describing the strong forces between quarks and gluons~\cite{Gross_2023}. Driven by rapid advances in high-performance computing and numerical techniques, lattice QCD simulations can now determine static properties of QCD -- such as the hadron mass spectrum, nuclear force, and the finite-temperature equation of state -- with unprecedented precision. These achievements have established QCD as a precision science, and further improvements are pursued on cutting-edge supercomputers, particularly in efforts to probe physics beyond the Standard Model~\cite{aoki2025flagreview2024}.

Despite the great success of lattice simulations, a long-standing challenge remains in computing the real-time dynamics of QCD, or more generally, of gauge theories on classical computers. In real-time simulations, the importance sampling of the path integral, which is the core of lattice simulations, does not work due to the complex phase of the action. This is known as the sign problem and severely limits the first-principles studies of several key phenomena in high-energy physics, such as the thermalization dynamics of QCD relevant to the formation of quark-gluon plasma in heavy-ion collisions, and the real-time dynamics of phase transitions in the early universe.
Note also that lattice simulations suffer from the sign problem even in static problems at finite baryon densities~\cite{deforcrand2010simulatingqcdfinitedensity}, hindering accurate determination of the QCD equation of state that is crucial for understanding the interior of neutron stars.

The rapid progress of quantum computing technologies has opened new avenues for simulating complex quantum many-body systems.
Motivated by these developments~\cite{Banuls:2019bmf,DiMeglio:2023nsa}, significant effort has been devoted to applying quantum computers to lattice gauge theories~\cite{Klco:2018kyo,Klco:2019evd,Ciavarella:2021nmj,Atas:2021ext,Hayata:2021kcp,deJong:2021wsd,Ciavarella:2021lel,Nguyen:2021hyk,Atas:2022dqm,ARahman:2022tkr,Charles:2023zbl,Farrell:2023fgd,Schuster:2023klj,Angelides:2023noe,Farrell:2024fit,Ciavarella:2024fzw,Hayata:2024smx,Cochran:2024rwe,Hayata2025,Crippa:2024hso,qr72-51v1,Gonz_lez_Cuadra_2025,PhysRevD.111.054501,mwy1-v9hk,davoudi2025quantumcomputationhadronscattering,ingoldby2025realtimescatteringquantumcomputers,cobos2025realtimedynamics21dgauge}. For example, the string breaking dynamics has been demonstrated on quantum computers. However, most existing studies focus on abelian gauge theories, such as the $Z_2$ gauge theory, or nonabelian gauge theories in $(1+1)$ dimensions. Although several works have explored nonabelian gauge theories in $(2+1)$ dimensions, their Hamiltonians are simplified such that the essential dynamics reduces to that of  abelian gauge theories or even that of spin models. To our knowledge, the real-time dynamics of a genuinely nonabelian gauge theory, retaining its intrinsic nonabelian structure, has yet to be simulated even for a simple toy model.

Hereafter, by nonabelian we mean that the composition rule of Wilson lines follows that of nonabelian anyons: the fusion of two Wilson lines does not form a (discrete) group, but does the fusion algebra. As detailed below, we realize such genuinely nonabelian dynamics by implementing Trotterized quantum circuits based on $F$-moves, which act as local unitary transformations between string networks.

In this paper, we demonstrate a digital quantum simulation of the thermalization dynamics of a nonabelian lattice gauge theory using a trapped-ion computer. A direct implementation of lattice QCD may be beyond the reach of present quantum devices because it requires both an excessive number of qubits and prohibitively deep quantum circuits. To alleviate this limitation, we introduce a simplified model that retains the essential nonabelian fusion rule of Wilson lines. Using the simplified model whose gauge dynamics is described by the simplest nonabelian anyons, known as the Fibonacci anyons, we study what is required to leverage the full potential of the present noisy quantum devices to perform digital quantum simulations of nonabelian lattice gauge theories.

A key challenge is the characterization of hardware- and circuit-dependent noise, which is essential for making quantitative predictions. Unlike in digital simulations of models such as the transverse-field Ising model, the dominant errors in our circuit arise not from two-qubit gates -- usually the primary source -- but from memory noise incurred while qubits wait for operations of the $F$-moves. We show that these memory errors can be substantially mitigated by dynamical decoupling. 

Our experiment demonstrates that current noisy quantum devices have the capability to simulate the nontrivial real-time dynamics of nonabelian gauge theories if quantum circuits are carefully designed and appropriate error-mitigation techniques are employed. In particular, identifying the dominant noise sources and selecting effective mitigation strategies are essential for scaling up such experiments and enabling studies of nonequilibrium dynamics in regimes that are classically intractable. We also emphasize that our simulation explicitly implements the $F$-moves, leveraging the strengths of digital quantum computation. Taken together, our results pave the way for quantum simulations of nonabelian gauge theories on noisy quantum hardware, moving toward addressing practical high-energy physics problems with quantum computers.

\section{\texorpdfstring{$q$}{q}-deformed Yang-Mills theory on a honeycomb lattice}
\label{sec:model}

We start with the $q$-deformed formulation of the lattice Yang-Mills theory on a honeycomb lattice~\cite{Levin:2004mi,Zache:2023dko,Hayata:2023puo,Hayata:2023bgh,Hayata:2026xeo} with the $F$-moves describing the fusion of Fibonacci anyons~\cite{Hayata:2023puo}. The Hamiltonian is essentially the same as that of a microscopic model of interacting anyons proposed in Ref.~\cite{Gils2009}. The basis of quantum states is represented by a network of Wilson lines, or equivalently, a network of strings~\cite{Levin:2004mi}. For instance, the basis of the quantum state of a one-plaquette system is graphically represented by
\begin{equation}
\label{eq:edges}
  \vcenter{\hbox{\includegraphics[scale=0.28]{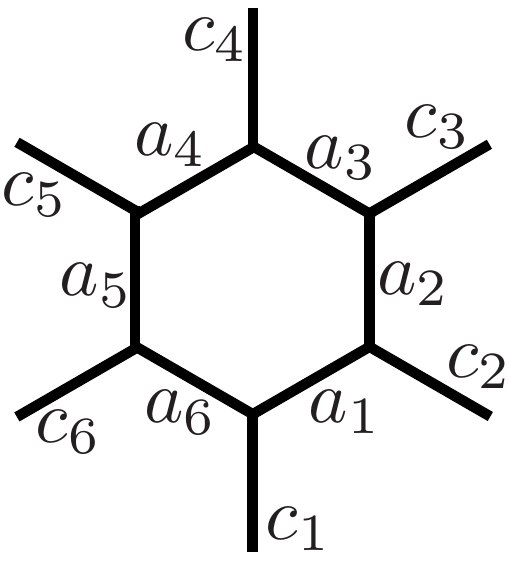}}}\quad,
\end{equation}
where edges are labeled by the Dynkin indices or the type of anyons excited on the edges. For instance, $a_i$ and $c_i$ take half-integer values ranging from $0$ to $k/2$ in the $q$-deformed $\mathrm{SU}(2)_k$ Yang-Mills theory, and the composition rule of Wilson lines is given as $a\times b=\sum_c N_{ab}^c c$ with
\begin{equation}
 N_{ab}^c\coloneqq \begin{cases}
  1 &
  j_{a}+j_{b}\geq j_{c},
  j_{b}+j_{c}\geq j_{a},
  j_{c}+j_{a}\geq j_{b},
  \\
  & j_{a}+j_{b}+j_{c}\in\mathbb{Z},\
   \text{and}\ j_a+j_b+j_c\leq k
  \\
  0 & \text{else}
  \end{cases}.
  \label{eq:fusion q}
  \end{equation}
Thus, the fusion of two Wilson lines does not form a group except $k=1$~\cite{Hayata:2026xeo}, but does the fusion algebra. From now on, we consider the so-called Fibonacci anyons as excitations, which correspond to the integer-spin sector of the $q$-deformed $\mathrm{SU}(2)_3$ Yang-Mills theory. Then, there are only two possible types: one is the trivial particle $1$, and the other is the Fibonacci anyon $\tau$. Thus, $a_i$ and $c_i$ take either $1$ or $\tau$, which are mapped onto qubits. In the language of $q$-deformed $\mathrm{SU}(2)_3$ Yang-Mills theory, $1$ and $\tau$ correspond to the spin-$0$ and spin-$1$ representations, respectively. The fusion rule of $1$ and $\tau$ is given as
\begin{align}
    \label{eq:fusion}
1\times \tau =\tau ,\qquad \tau\times 1 =\tau ,\qquad \tau \times \tau =1+\tau.
\end{align}
All string networks are not necessarily physically relevant states, and the logical Hilbert space is described by a set of them that satisfy the constraints at the vertices. The local basis satisfying the constraints by the fusion rule~\eqref{eq:fusion} at a vertex is graphically represented as
\begin{align}
    \label{eq:gauss}
    \vcenter{\hbox{\includegraphics[scale=0.2]{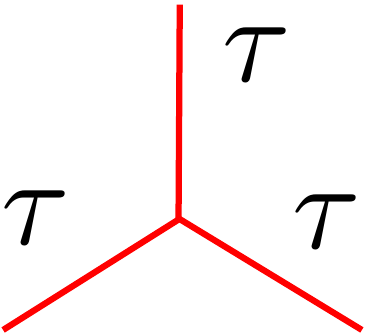}}}\quad,\qquad
    \vcenter{\hbox{\includegraphics[scale=0.2]{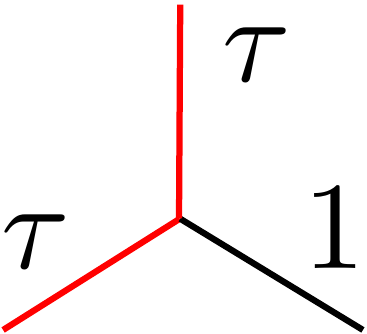}}}\quad,\qquad
    \vcenter{\hbox{\includegraphics[scale=0.2]{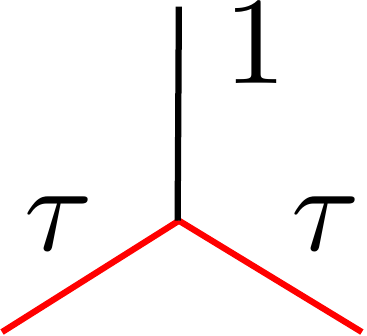}}}\quad,\notag \\
    \vcenter{\hbox{\includegraphics[scale=0.2]{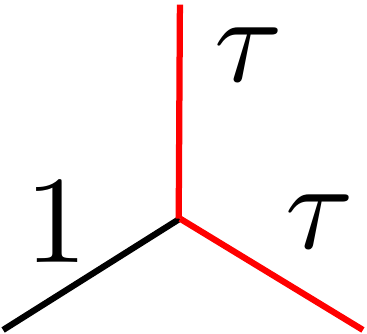}}}\quad,\qquad
    \vcenter{\hbox{\includegraphics[scale=0.2]{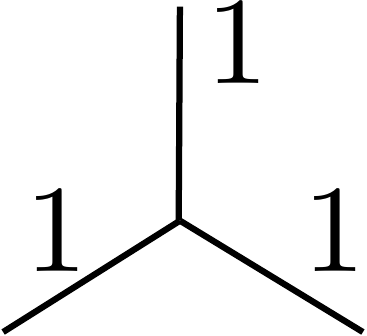}}}\quad.
\end{align}
Notice that the annihilation of a single $\tau$ anyon on a vertex is not allowed. Those constraints are nothing but the Gauss-law constraints in lattice gauge theories. The entire basis of the logical Hilbert space is constructed by connecting the local basis satisfying~\eqref{eq:gauss} to form a honeycomb lattice. Notice that the wave function identically satisfies the constraints under the Hamiltonian evolution if the initial state does.

We here introduce the Hamiltonian describing the dynamics of our system. The Kogut-Susskind Hamiltonian of the lattice Yang-Mills theory is
\begin{equation}
\begin{split}
  H_\mathrm{YM} &= H_E+H_M
  \\
  &=\frac{1}{2}\sum_{e\in {\mathcal{E}}} E_i^2(e)
  - K \sum_{f\in \mathcal{F}} \tr U_{\tau} (f) ,
\end{split}
  \label{eq:YM}
\end{equation}
where $\mathcal{E}$ and $\mathcal{F}$ are the sets of edges and faces of a honeycomb lattice, respectively~\cite{PhysRevD.11.395}. $E_i^2(e)$ and $\tr U_{\tau} (f)$ are the squares of the electric-field operators and the plaquette operator, respectively. In the string-net basis, they are graphically represented as
\begin{equation}
  E_i^2 \vcenter{\hbox{\includegraphics[scale=0.3]{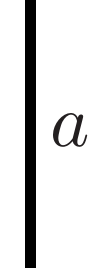}}}
  = \delta_{a\tau}
  \vcenter{\hbox{\includegraphics[scale=0.3]{a.pdf}}},
  \label{eq:action_electric_field}
\end{equation}
and
\begin{equation}
\label{eq:action_trU_general}
\begin{split}
     & \tr U_{\tau} \quad
       \vcenter{\hbox{\includegraphics[scale=0.22]{plaquetteAction0.pdf}}}
     =\prod_{i=1}^{6} \sum_{a'_{i}}
       [F_{a'_{i}}^{c_{i}a_{i-1}\tau}]_{a_{i}{a'}_{i-1}}
         \vcenter{\hbox{\includegraphics[scale=0.22]{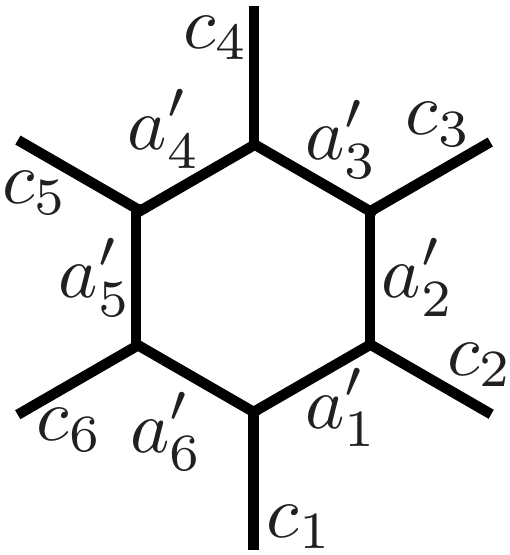}}}\,,
\end{split}
\end{equation}
where $a'_0=a'_6$, and $[F^{abc}_d]_{ef}$ is the $F$-move, which describes the local unitary transformation between string networks and is defined as
\begin{align}
  \ \vcenter{\hbox{\includegraphics[scale=0.18]{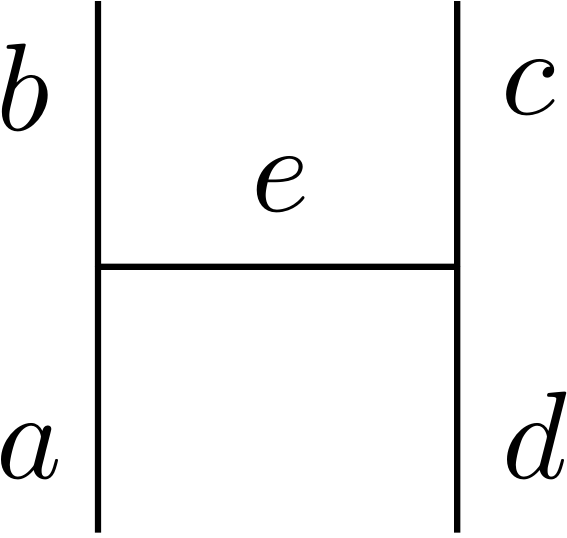}}}
&= \sum_{f} [F^{abc}_d]_{ef}
\ \vcenter{\hbox{\includegraphics[scale=0.18]{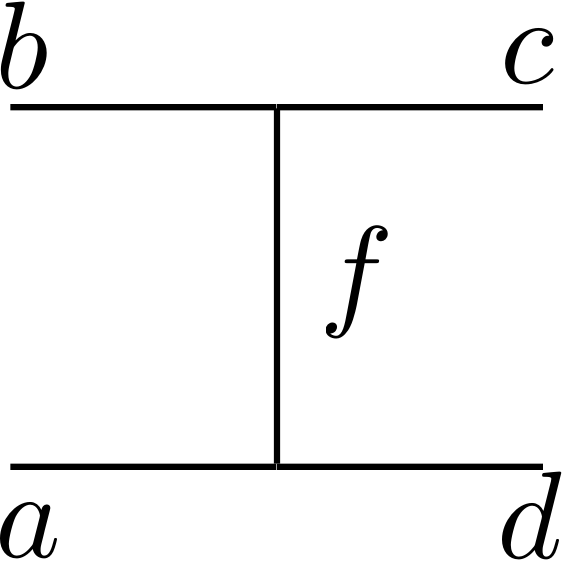}}} \;\;.
\label{eq:Fmove}
\end{align}
The nontrivial $F$-move of the Fibonacci anyons is 
\begin{equation}
[F^{\tau\tau\tau}_\tau]=
    \begin{pmatrix}
    \frac{1}{\varphi} & \frac{1}{\varphi^{\frac{1}{2}}} \\
    \frac{1}{\varphi^\frac{1}{2}} & -\frac{1}{\varphi} 
    \end{pmatrix} ,
    \label{eq:nontrivialF}
\end{equation}
with $\varphi$ being the golden ratio, $\varphi=({1+\sqrt{5}})/{2}$, and the others are $[F^{abc}_d]_{e,f}=1$ if the labels in Eq.~\eqref{eq:Fmove} satisfy the constraints~\eqref{eq:gauss} at the vertices and otherwise $[F^{abc}_d]_{e,f}=0$. The last property of the $F$-moves guarantees that the Hamiltonian commutes with the constraints.

In principle, the Trotterized time evolution generated by the Hamiltonian~\eqref{eq:YM} can be simulated on digital quantum computers using the circuit representation of the $F$-moves as described in Ref.~\cite{Zache:2023dko}. However, each plaquette term requires ten $F$-moves per Trotter step~\cite{Zache:2023dko}, and the implementation of the $F$-moves is very challenging on current noisy quantum devices (the explicit circuit will be given shortly). As a result, the Hamiltonian~\eqref{eq:YM} involves an excessive number of two-qubit gates in one Trotter step, leaving the Trotter evolution (even the Floquet evolution with a large Trotter step) infeasible on present quantum devices. Indeed, the explicit action of the $F$-moves can be avoided using the dual Hamiltonian if gauge fields are effectively abelian~\cite{Hayata2025}. Such a dual Hamiltonian is no longer useful to avoid the $F$-moves for our case, and we need to face them directly. Hence, in this paper, we modify the Hamiltonian to reduce the cost of the $F$-moves. 

Our prescription is very simple: we apply the plaquette interactions and the electric terms only to a subset of the graph as described below. Since we can understand the plaquette interaction as a penalty term that does not allow the loop excitation~\cite{Gils2009}, this prescription means a (strong) perturbation that induces loop excitations in terms of the original honeycomb model. Importantly, this modification does not break the gauge symmetry at all. In graphical representations, the plaquette terms are represented as holes~\cite{Hayata:2023bgh,Gils2009}, and we remove some holes as follows:
\begin{equation}
\begin{split}\label{eq:original_lattice}
    \vcenter{\hbox{\includegraphics[scale=0.2]{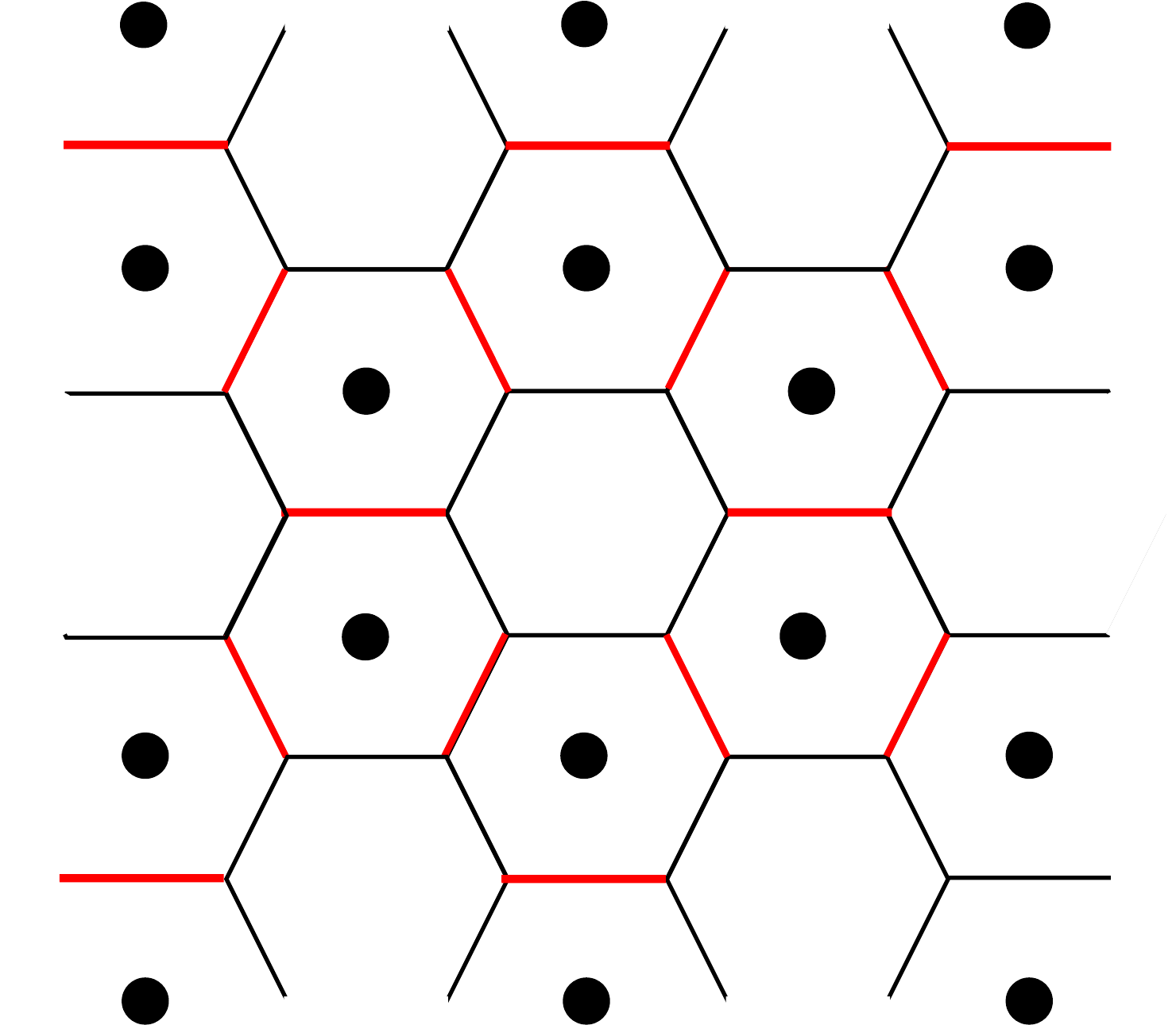}}} \;\;\;\;,
\end{split}
\end{equation}
where we apply the $E^2$ operator to the red edges (that correspond to tubes in Ref.~\cite{Gils2009}), and the $\tr U_{\tau}(f)$ operator to plaquettes with holes. Notice that the resulting system is topologically equivalent to the theory on a triangular lattice. This provides an alternative perspective, showing that our prescription preserves the gauge symmetry. Under the prescription, we can move to the topologically equivalent graph, i.e., the decorated honeycomb lattice, following Refs.~\cite{Gils2009,Hayata2025}:
\begin{equation}
\begin{split}
    \vcenter{\hbox{\includegraphics[scale=0.25]{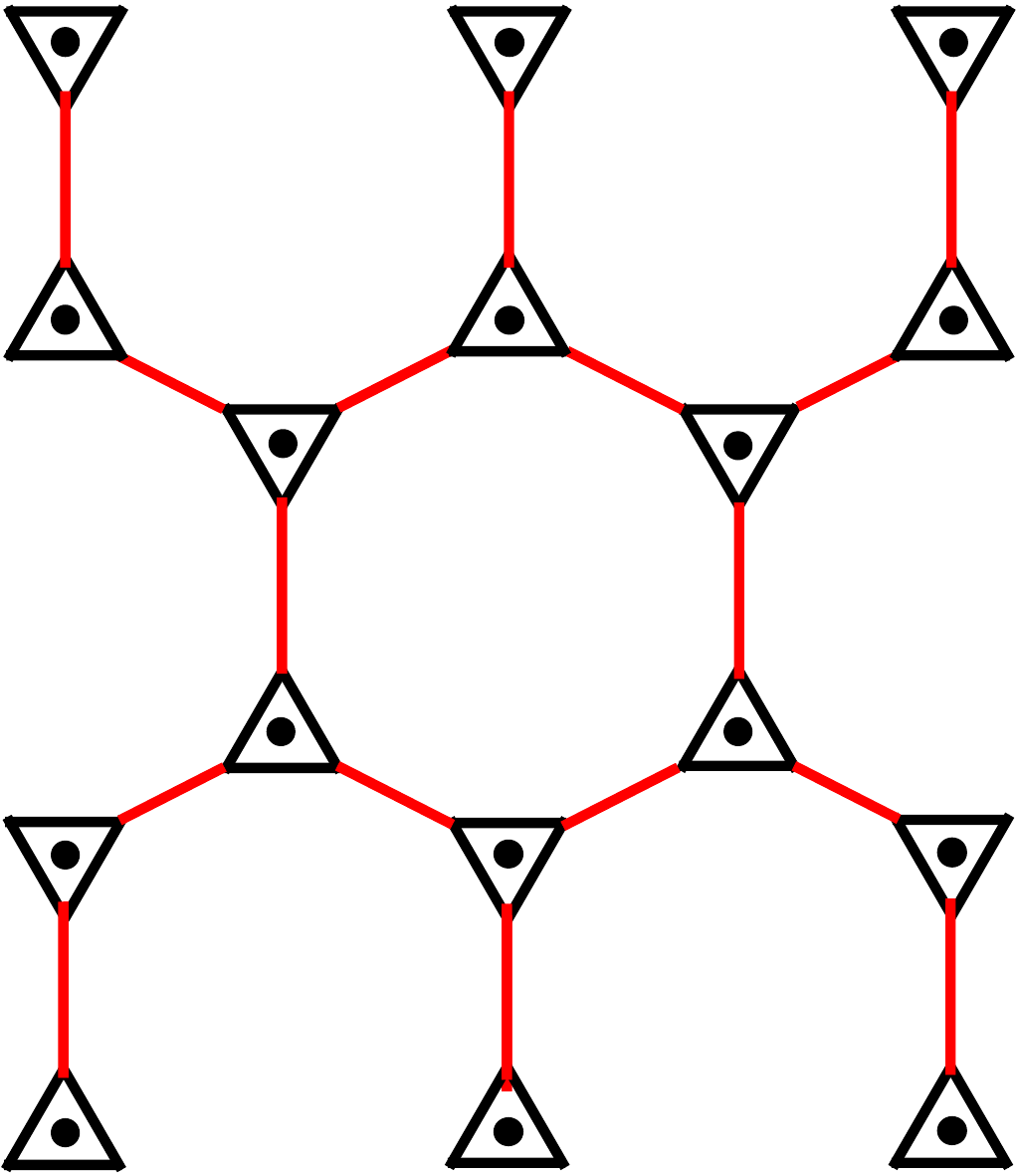}}} \;\;\;\;,
\end{split}
\end{equation}
where we applied the $F$-moves to the red edges successively. In this basis, the $E^2$ operator acts as
\begin{equation}
\begin{split}
       E^2 \;
       \vcenter{\hbox{\includegraphics[scale=0.25]{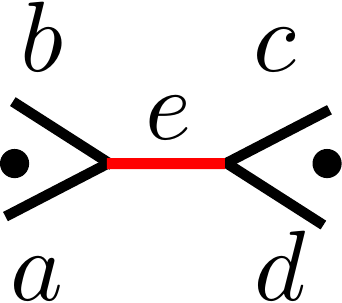}}}
    &=\sum_f [F_{d}^{abc}]_{ef}
       E^2 \;
         \vcenter{\hbox{\includegraphics[scale=0.25]{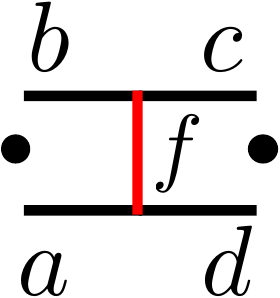}}}
    \\
       &=\sum_f[F_{d}^{abc}]_{ef}
       \delta_{f\tau} \;
         \vcenter{\hbox{\includegraphics[scale=0.25]{outE1.pdf}}}
    \\
       &=\sum_{f,e^\prime}[F_{d}^{abc}]_{ef}[F_{d}^{abc}]_{e^\prime f}
       \delta_{f\tau} \;
         \vcenter{\hbox{\includegraphics[scale=0.25]{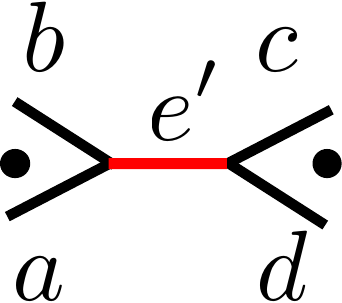}}} \;\;.
         \label{eq:action_E2}
\end{split}
\end{equation}
Notice that the $E^2$ operator is a diagonal operator in the original honeycomb graph~\eqref{eq:original_lattice}.
Similarly, its exponential reads
\begin{equation}
\begin{split}
       \e^{-\im dt E^2} \;
       \vcenter{\hbox{\includegraphics[scale=0.25]{inE1.pdf}}}\;\;
       &=\sum_{f,e^\prime}[F_{d}^{abc}]_{ef}[F_{d}^{abc}]_{e^\prime f}
       \e^{-\im dt \delta_{f\tau}} \;
         \vcenter{\hbox{\includegraphics[scale=0.25]{outE1_2.pdf}}} \;\;,
         \label{eq:action_E2_exp}
\end{split}
\end{equation}
where $\e^{-\im dt \delta_{f\tau}}$ can be implemented as the $R_z$ single-qubit rotation gate in qubit quantum devices, and this uses two $F$-moves in one Trotter step. We map the $1$ and $\tau$ particles to the $0$ and $1$ states in qubits, respectively. Then, $\e^{-\im dt \delta_{f\tau}}$ can be implemented as $R_z(-dt)=\e^{\im \frac{dt}{2}Z}$ up to the irrelevant global phase.

Next, the plaquette operator acts as
\begin{equation}
\begin{split}
       & \tr U_{\tau} \;
       \vcenter{\hbox{\includegraphics[scale=0.25]{in2.pdf}}}
       \\
      &=
       \sum_{b_3}[F_{a_2}^{a_1c_1c_3}]_{a_3b_3}
         \tr U_{\tau} \;\vcenter{\hbox{\includegraphics[scale=0.25]{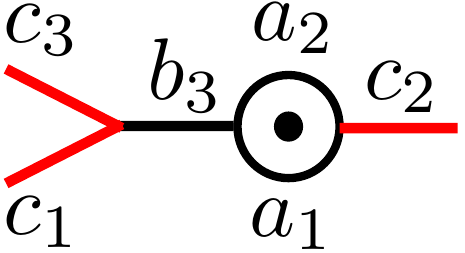}}}
         \\
                &=
       \sum_{b_2,b_3}[F_{a_2}^{a_1c_1c_3}]_{a_3b_3}[F_{a_1}^{a_1b_3c_2}]_{a_2b_2}
         \tr U_{\tau} \;\vcenter{\hbox{\includegraphics[scale=0.25]{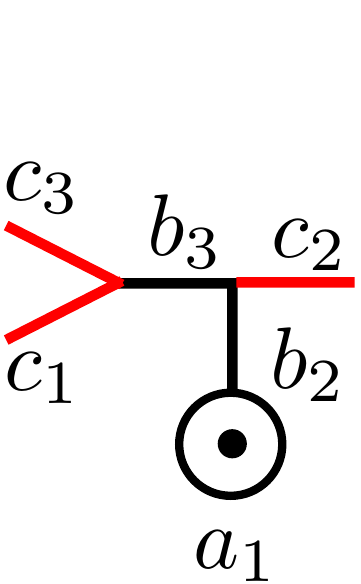}}}
        \\
                &=
       \sum_{b_1,b_2,b_3}[F_{a_2}^{a_1c_1c_3}]_{a_3b_3}[F_{a_1}^{a_1b_3c_2}]_{a_2b_2}[S^{b_2}]_{a_1b_1}
         \;\vcenter{\hbox{\includegraphics[scale=0.25]{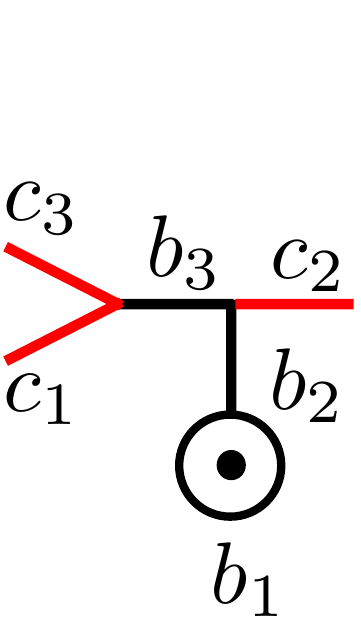}}}
        \\
                &=
       \sum_{b_1,b_2,b_3,b_2^\prime}[F_{a_2}^{a_1c_1c_3}]_{a_3b_3}[F_{a_1}^{a_1b_3c_2}]_{a_2b_2}[S^{b_2}]_{a_1b_1}
         \;\\
         &\hspace{10em}\times [F_{b_1}^{b_1b_3c_2}]_{b_2^\prime b_2}\vcenter{\hbox{\includegraphics[scale=0.25]{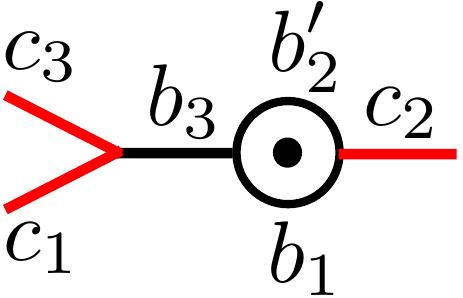}}}
        \\
                &=
       \sum_{b_1,b_2,b_3,b_2^\prime,b_3^\prime}[F_{a_2}^{a_1c_1c_3}]_{a_3b_3}[F_{a_1}^{a_1b_3c_2}]_{a_2b_2}[S^{b_2}]_{a_1b_1}
       \\
       & \hspace{7em}
       \times[F_{b_1}^{b_1b_3c_2}]_{b_2^\prime b_2}[F_{b_2^\prime}^{b_1c_1c_3}]_{b_3^\prime b_3}
         \vcenter{\hbox{\includegraphics[scale=0.25]{out2_5.pdf}}}
         ,
         \label{eq:action_trU}
\end{split}
\end{equation}
where we defined the action of $\tr U_{\tau} $ to a tadpole graph as
\begin{equation}
\tr U_{\tau} \;
       \vcenter{\hbox{\includegraphics[scale=0.25]{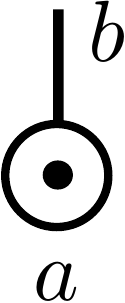}}}
       =\sum_{a^\prime}[S^b]_{aa^\prime} \;\vcenter{\hbox{\includegraphics[scale=0.25]{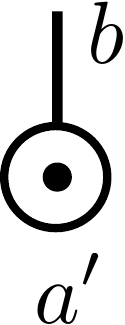}}} \,.
\end{equation}
Taking $a_i=a$ for $i=1,2,\cdots,6$, $c_1=b$, and $c_i=1$ for $i=2,3,\cdots,6$ in Eq.~\eqref{eq:action_trU_general}, we obtain $[S^b]_{aa^\prime}=[F_{a'}^{ba\tau}]_{aa'}$. A tadpole graph must satisfy the constraints~\eqref{eq:gauss}. Thus, if $b=\tau$, the vertex constraint requires $a=a^\prime=\tau$, and $[S^\tau]_{\tau\tau}=[F_{\tau}^{\tau\tau\tau}]_{\tau\tau}=-1/\varphi$. It then follows that $[S^\tau]_{aa^\prime}=-\delta_{a\tau}\delta_{a^\prime\tau}/\varphi$. Next, for $b=1$, we have $[S^1]_{aa^\prime}=[F_{a^\prime}^{1a\tau}]_{aa^\prime}$, which equals unity if the labels in Eq.~\eqref{eq:Fmove} satisfy the constraints~\eqref{eq:gauss} at the vertices, and vanishes otherwise.
Consequently, $[S^{1}]_{aa^\prime}$ takes the form
\begin{equation}
[S^{1}]=
    \begin{pmatrix}
    0 & 1 \\
    1 & 1 
    \end{pmatrix} .
\end{equation}

Similarly, $\e^{-\im dt\tr U_{\tau}}$ can be implemented by replacing $S$ with $\e^{-i dt S}$, which is a controlled unitary gate in the quantum circuit. $\e^{-\im dt\tr U_{\tau}}$ consumes two general $F$-moves and two special $F$-moves in which two of four arguments of the $F$-moves are the same. Notice that $\e^{-\im dt \tr U_{\tau}}$ consumes eight general $F$-moves and two special $F$-moves (ten $F$-moves in total for each plaquette term) in the original honeycomb graph~\cite{Zache:2023dko}.

\section{Methods}

\subsection{Model}
\begin{figure}[t]
  \includegraphics[width=.35\textwidth]
  {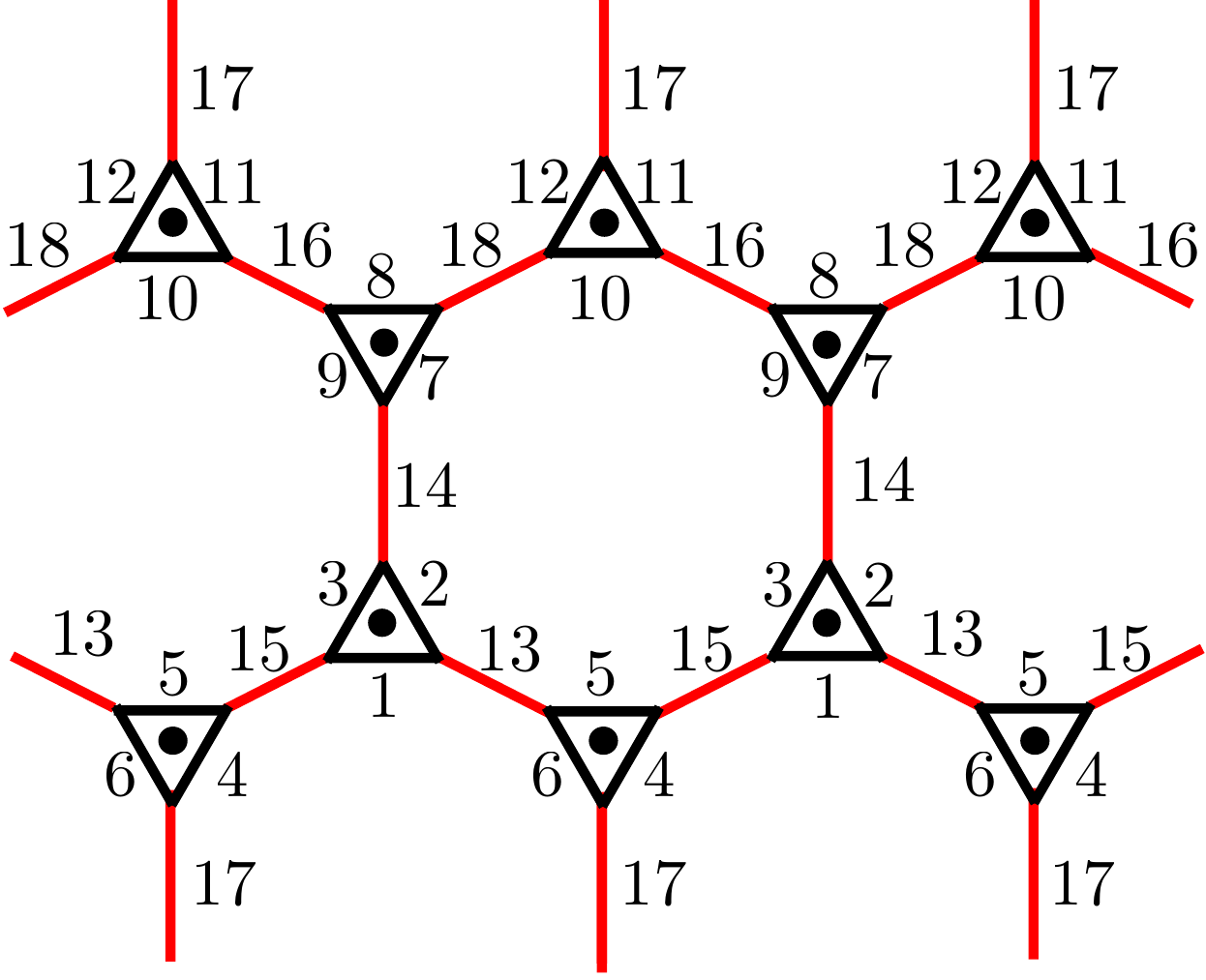}
  \caption{
    \label{fig:system}
    Qubit layout simulated in experiments. 
    }
\end{figure}%
We simulate the thermalization dynamics of an 18 qubit system shown in Fig.~\ref{fig:system}. We consider the periodic boundary conditions.
In the Hamiltonian, we have four plaquette (magnetic) terms that act on the following qubits:
\begin{equation}
\label{eq:faces}
\begin{split}
       \vcenter{\hbox{\includegraphics[scale=0.3]{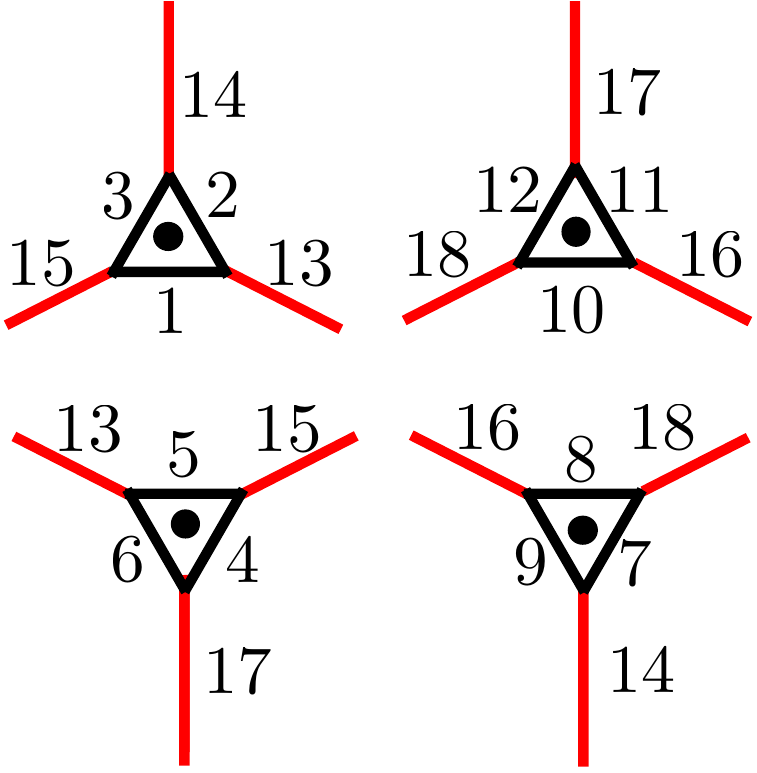}}} \hspace{2em},
\end{split}
\end{equation}
and six electric terms that act on the following qubits:
\begin{equation}
\label{eq:edges2}
\begin{split}
       \vcenter{\hbox{\includegraphics[scale=0.3]{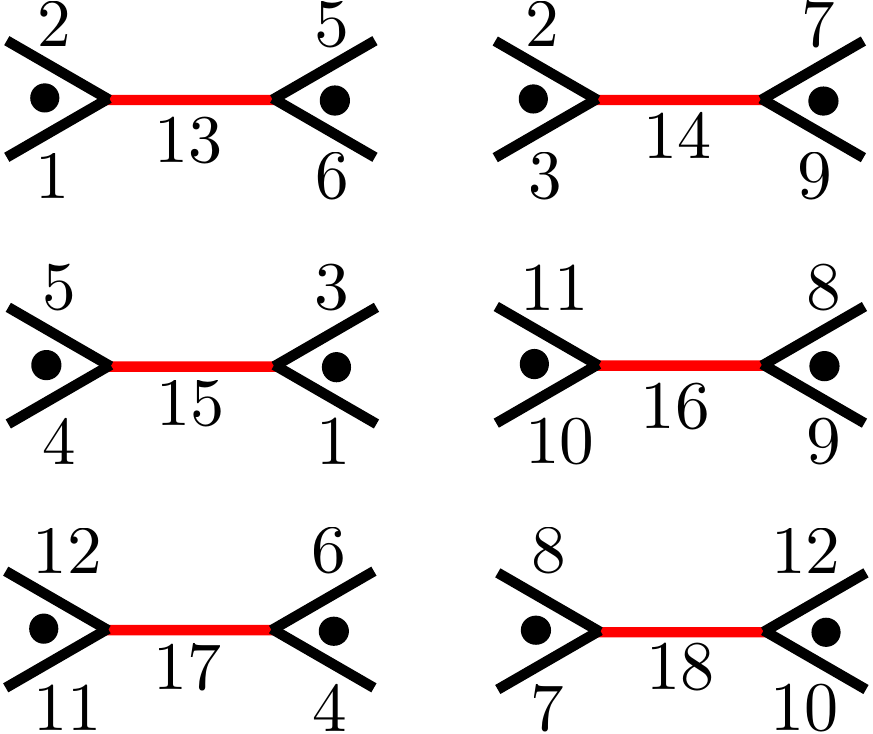}}} \hspace{2em}.
\end{split}
\end{equation}
This may be a minimal (2+1)-dimensional system.

We consider the 2nd order Trotter decomposition of the Hamiltonian evolution
\begin{eqnarray}
    \e^{-iHt} &\approx& \left(\e^{-\im \frac{dt}{2} H_M}\e^{-\im dt H_E}\e^{-\im \frac{dt}{2} H_M}\right)^N 
    \notag \\
    &=&\e^{-\im \frac{dt}{2} H_M}\left(\e^{-\im dt H_E}\e^{-\im dt H_M}\right)^{N-1} \e^{-\im dt H_E}\e^{-\im \frac{dt}{2} H_M} ,
    \label{eq:1stTrotter}
\end{eqnarray}
where $t$, $N$, and $dt=t/N$ are the time interval, the number of Trotter steps, and the Trotter step, respectively. Since we measure the Wilson loop in our experiment, the last half-step $\e^{-\im \frac{dt}{2} H_M}$, which commutes with the Wilson loop operator, can be removed.

For a later purpose, we introduce the self-unitary Wilson loop as follows. First, the normalized regular Wilson loop 
\begin{equation}
    \tr U_\mathrm{reg}=\frac{1}{1+\varphi^2}\left(1+\varphi\; \tr U_\tau\right)
\end{equation}
is a projector, i.e., $(\tr U_\mathrm{reg})^2=\tr U_\mathrm{reg}$. Therefore, 
\begin{equation}
\tr U_\mathrm{P}=2\tr U_\mathrm{reg}-1 =\frac{1-\varphi^2}{1+\varphi^2}+\frac{2\varphi}{1+\varphi^2} \tr U_\tau,
\end{equation}
satisfies $(\tr U_\mathrm{P})^2=(2\tr U_\mathrm{reg}-1)^2=1$. We also define $S_\mathrm{P}$ as
\begin{equation}\label{eq:S_P}
S_\mathrm{P} =\frac{1-\varphi^2}{1+\varphi^2}+\frac{2\varphi}{1+\varphi^2} S .
\end{equation}
Its eigenvalues are $\pm1$, so that $S_\mathrm{P}$ and its exponential can be written as the Pauli $Z$ and phase rotation gate, respectively. This property is useful when we implement the Trotter evolution gates on real devices. Notice that ${2\varphi}/(1+\varphi^2)={2}/{\sqrt{5}}$, and $\e^{-\im Kdt \tr U_\tau}=\e^{-\im \frac{\sqrt{5}}{2}Kdt \tr U_\mathrm{P}}$ up to the irrelevant global phase.

\subsection{Gate decomposition}
We follow Refs.~\cite{Bonesteel:2012pkv,Xu:2024guv} for the circuit implementation of the $F$-moves.
An $F$-move can be represented by the 4-controlled gate operations,
\begin{equation}
\label{eq:fmove_gate}
    [F^{abc}_d] 
    \quad=\quad
    \begin{quantikz}[column sep=.1cm, row sep=.2cm]
         \lstick{$a$} 
         & \ctrl{1}
         &  
         & \ctrl{2}
         & 
         & \ctrl{2}
         & 
         & 
         \\
         \lstick{$b$} 
         & \ctrl{1}
         & \gate{X} 
         & 
         & \ctrl{1}
         & 
         & \gate{X} 
         & 
         \\
         \lstick{$c$} 
         & \ctrl{1}
         & 
         & \gate{X}
         & \ctrl{2}
         & \gate{X}
         & 
         & 
         \\
         \lstick{$d$} 
         & \ctrl{1}
         & \ctrl{-2} 
         &  
         & 
         & 
         & \ctrl{-2} 
         & 
         \\
         & \gate{F}
         &  
         & 
         & \gate{X}
         & 
         & 
         & 
    \end{quantikz}       \;\;,
\end{equation}
where the 4-controlled gate denoted by $F$ is the unitary gate of the nontrivial $F$-move $[F^{\tau\tau\tau}_\tau]$ in Eq.~\eqref{eq:nontrivialF}, and may be represented as
\begin{equation}
\begin{split}
    \begin{quantikz}[column sep=.1cm, row sep=.2cm]
         & \ctrl{1}
         & 
         \\
         & \ctrl{1}
         & 
         \\
         & \ctrl{1}
         & 
         \\
         & \ctrl{1}
         &
         \\
         & \gate{F}
         & 
    \end{quantikz}       
    \quad = \quad
    \begin{quantikz}[column sep=.1cm, row sep=.2cm]
         &
         & \ctrl{1}
         & 
         & 
         \\
         &
         & \ctrl{1}
         & 
         & 
         \\
         &
         & \ctrl{1}
         &
         & 
         \\
         &
         & \ctrl{1}
         &
         & 
         \\
         & \gate{R_y[\theta]}
         & \gate{X}
         & \gate{R_y[-\theta]}
         &  
    \end{quantikz}       ,
\end{split} 
\end{equation}
with $\theta=\arctan \varphi^{-\frac{1}{2}}$.
Using two ancilla qubits, we decompose the 5-qubit Toffoli (CCCCX) gate to the Toffoli (CCX) gate as 
\begin{equation}
    \begin{quantikz}[column sep=.1cm, row sep=.2cm]
         & \ctrl{1}
         & 
         \\
         & \ctrl{1}
         & 
         \\
         & \ctrl{1}
         & 
         \\
         & \ctrl{1}
         &
         \\
         & \gate{X}
         & 
    \end{quantikz}
    \quad = \quad
    \begin{quantikz}[column sep=.1cm, row sep=.2cm]
         & \ctrl{1}
         & 
         & 
         & 
         & \ctrl{1}
         & 
         \\
         & \ctrl{1}
         & 
         & 
         & 
         & \ctrl{1}
         & 
         \\
         \lstick{$|0\rangle$}
         & \gate{X}
         & \ctrl{1}
         & 
         & \ctrl{1}
         & \gate{X}
         & 
         \rstick{$|0\rangle$}
         \\
         &
         & \ctrl{1}
         & 
         & \ctrl{1}
         & 
         & 
         \\
        \lstick{$|0\rangle$}
         &
         & \gate{X}
         & \ctrl{1}
         & \gate{X}
         & 
         & 
         \rstick{$|0\rangle$}
         \\
         &
         & 
         & \ctrl{1}
         & 
         & 
         & 
         \\
         & 
         & 
         & \gate{X}
         & 
         & 
         &  
    \end{quantikz}       ,
\end{equation}
where the ancilla qubits are the qubits whose initial and final states are $|0\rangle$.
We also need the $F$-moves such that two of four arguments (control qubits) are the same, i.e., $[F_a^{abc}]$.
Such $F$-moves may be represented as
\begin{equation}
\label{eq:fmove_gate_3}
    [F^{abc}_a] 
    \quad=\quad
    \begin{quantikz}[column sep=.1cm, row sep=.2cm]
         \lstick{$a$} 
         & \ctrl{1}
         &  \ctrl{1}
         & \ctrl{2}
         & 
         & \ctrl{2}
         & \ctrl{1}
         & 
         \\
         \lstick{$b$} 
         & \ctrl{1}
         & \gate{X} 
         & 
         & \ctrl{1}
         & 
         & \gate{X} 
         & 
         \\
         \lstick{$c$} 
         & \ctrl{1}
         & 
         & \gate{X}
         & \ctrl{1}
         & \gate{X}
         & 
         & 
         \\
         & \gate{F}
         &  
         & 
         & \gate{X}
         & 
         & 
         & 
    \end{quantikz}       .
\end{equation}
This can be obtained by assuming $a=d$ in the $F$-moves in Eq.~\eqref{eq:fmove_gate}.
The 3-controlled $F$ is defined by
\begin{equation}
    \begin{quantikz}[column sep=.1cm, row sep=.2cm]
         & \ctrl{1}
         & 
         \\
         & \ctrl{1}
         & 
         \\
         & \ctrl{1}
         &
         \\
         & \gate{F}
         & 
    \end{quantikz}       
    \quad = \quad
    \begin{quantikz}[column sep=.1cm, row sep=.2cm]
         &
         & \ctrl{1}
         & 
         & 
         \\
         &
         & \ctrl{1}
         &
         & 
         \\
         &
         & \ctrl{1}
         &
         & 
         \\
         & \gate{R_y[\theta]}
         & \gate{X}
         & \gate{R_y[-\theta]}
         &  
    \end{quantikz} ,
\end{equation}
and we can decompose the 4-qubit Toffoli (CCCX) gate to the Toffoli (CCX) gate, using one ancilla qubit,  as 
\begin{equation}
    \begin{quantikz}[column sep=.1cm, row sep=.2cm]
         & \ctrl{1}
         & 
         \\
         & \ctrl{1}
         & 
         \\
         & \ctrl{1}
         &
         \\
         & \gate{X}
         & 
    \end{quantikz}       
    \quad = \quad
    \begin{quantikz}[column sep=.1cm, row sep=.2cm]
         & \ctrl{1}
         & 
         & \ctrl{1}
         & 
         \\
         & \ctrl{1}
         & 
         & \ctrl{1}
         & 
         \\
         \lstick{$|0\rangle$}
         & \gate{X}
         & \ctrl{1}
         & \gate{X}
         & 
         \rstick{$|0\rangle$}
         \\
         &
         & \ctrl{1}
         & 
         & 
         \\
         & 
         & \gate{X}
         & 
         &  
    \end{quantikz}       .
\end{equation}
Finally, we need $\e^{-\im \frac{dt}{2} S_\mathrm{P}}$, which may be written as
\begin{equation}
    \e^{-\im \frac{dt}{2} S_\mathrm{P}} 
    =
    \begin{quantikz}[column sep=.1cm, row sep=.2cm]
         & \gate{X}
         & 
         & \ctrl{1}
         & 
         & \gate{X}
         & \ctrl{1}
         & 
         \\
         & \gate{Z}
         & \gate{R_y[\bar{\theta]}}
         & \gate{R_z[dt]}
         & \gate{Z}
         & \gate{R_y[\bar{\theta]}}
         & \gate{R_z[dt]}
         & 
    \end{quantikz}       \;,
\end{equation}
where $\bar{\theta}=2\arctan\varphi$.
Notice that
\begin{equation}
\begin{split}
    [S^{1}_\mathrm{P}]&=
    \begin{pmatrix}
    \frac{1-\varphi^2}{1+\varphi^2} & \frac{2\varphi}{1+\varphi^2} \\
    \frac{2\varphi}{1+\varphi^2} & \frac{1+2\varphi-\varphi^2}{1+\varphi^2} 
    \end{pmatrix}
\\
    &=\frac{1}{\sqrt{1+\varphi^2}}\begin{pmatrix}
    1 & \varphi \\
    \varphi & -1 
    \end{pmatrix}\begin{pmatrix}
    1 & 0 \\
    0 & -1 
    \end{pmatrix}
    \frac{1}{\sqrt{1+\varphi^2}}
    \begin{pmatrix}
    1 & \varphi \\
    \varphi & -1 
    \end{pmatrix}
    \\
    &=R_y(\bar{\theta})Z Z R_y(\bar{\theta})Z
\end{split} \;,
\end{equation}
and  
\begin{equation}
\begin{split}
    [S^{\tau}_\mathrm{P}]&=
    \begin{pmatrix}
    1 & 0 \\
    0 & -1 
    \end{pmatrix}
\end{split} \;.
\end{equation}
If the control anyon is $\tau$ (1 in qubits), the target anyon (qubit) is always $\tau$ (1). Thus, $[S^{\tau}_\mathrm{P}]_{11}$ can be arbitrary since it does not act on the logical Hilbert space. Although the actual value of $[S^{\tau}_\mathrm{P}]_{11}$ in Eq.~\eqref{eq:S_P} is $-1/\sqrt{5}$, we choose $[S^{\tau}_\mathrm{P}]_{11}=1$, so as $[S^{\tau}_\mathrm{P}]$ to be implemented as the Pauli $Z$. Using those gates, we can simulate the Trotter circuit of our model. For additional gates that we need to measure the Wilson loop, see Appendix~\ref{app:Wilson}.

We remark that some Toffoli gates in $F$-moves whose target qubits are ancilla qubits can be replaced with the relative-phase Toffoli gates~\cite{song2003,Maslov2016}. The latter can be implemented more efficiently than the Toffoli gate. This is apparent since the ancilla qubits are in $|0\rangle$, and the phase does not appear~\cite{Maslov2016}. Also, we do not apply the (relative-phase) Toffoli gates to restore ancilla qubits to their initial state $|0\rangle$, but reset ancilla qubits after applying the Toffoli gate to the target qubit of the $F$-move. Those tricks reduce the cost of two-qubit gates in the $F$-moves almost by half.

\begin{figure}[!t]
  \includegraphics[width=.49\textwidth]{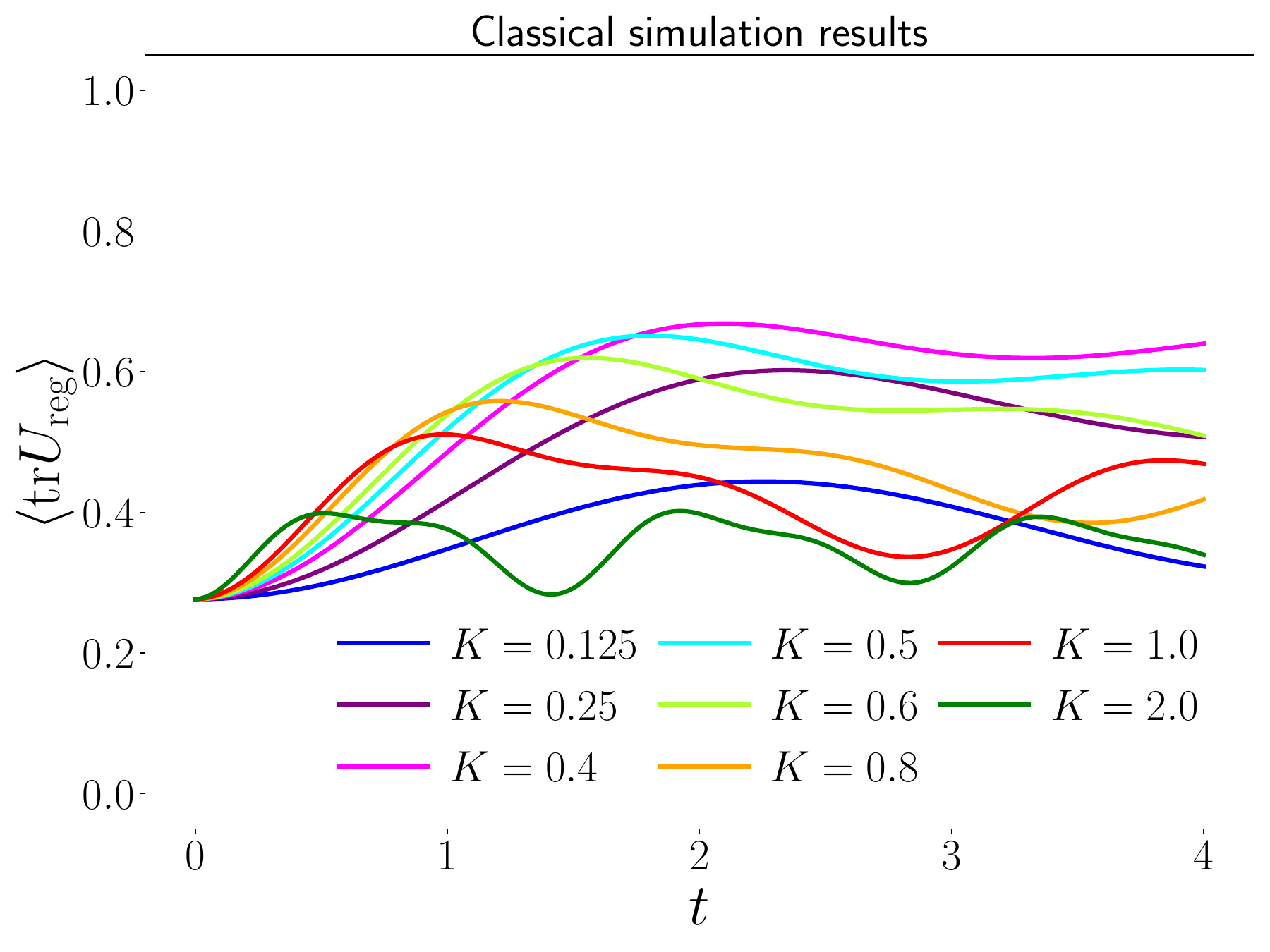}
  \caption{
    \label{fig:Trotter_classical}
    Classical noiseless simulation of the $K$ dependence of the thermalization dynamics. The fast thermalization is realized at $K=0.5$ around $t=3.0$.
    }
\end{figure}

The $K$ dependence of the thermalization dynamics is shown in Fig.~\ref{fig:Trotter_classical}. We here classically simulate the 2nd order Trotter circuit~\eqref{eq:1stTrotter} with $dt=0.02$, and compute the expectation value of the regular Wilson loop operator $\tr U_\mathrm{reg}$. We choose the vacuum in the strong coupling limit $|\bm{0}\rangle\equiv|0,\cdots,0\rangle$ (in qubit representation) as the initial state. The Trotter step size is set sufficiently small so that the associated error becomes negligible. We take this classical simulation as the exact result in the following experiments. We see fast thermalization realized around $K=0.5$. Accordingly, we focus on the quantum simulation of the thermalization dynamics at $K=0.5$.

\section{Experiments}
\label{sec:experiments}

We experimentally demonstrate a digital quantum simulation of the nonabelian gauge theory on Quantinuum's H2-1 trapped-ion quantum computer~\cite{Moses2023, H2-1_spec}. The H2-1 is a programmable digital quantum computer that consists of 56 qubits and supports single-qubit rotation gates and two-qubit ZZPhase gates $\e^{-\im\frac{\theta}{2}Z\otimes Z}$ with an arbitrary rotation angle~$\theta$ as native gates. The two-qubit native gates can be applied to any pair of qubits, realizing all-to-all connectivity. The average single and two-qubit gate infidelities are about $0.003\%$ and $0.1\%$, respectively. State preparation and measurement errors are $0.1\%$ on average. See Refs.~\cite{Moses2023, H2-1_spec} for more details. Furthermore, dynamical decoupling (DD) is available on the H2 system. DD is an error suppression technique that inserts a sequence of single-qubit pulses on idling qubits to reduce the accumulation of coherent noise.

We find that a naive implementation of $\e^{-\im dt H_E}\e^{-\im dt H_M}$ uses $680$ ZZPhase gates at each Trotter step after compiling circuits with TKET. To reduce the gate count, we explicitly decompose the gates of the first Trotter step, which can be further optimized by taking the initial state into consideration. For details, see Appendix~\ref{app:first_step}. As a result, the number of ZZPhase gates in the first Trotter step is cut down to only $50$ ZZPhase gates, including the $F$-moves used for measurement. We also adopt CPflow~\cite{Nemkov:2022mah} to variationally optimize the circuit for $\e^{-\im dt H_E}\e^{-\im dt H_M}$ allowing approximation error~\cite{Xu:2024guv}. See Appendix~\ref{app:CPFlow} for details. Combining all the circuit optimization strategies together, the total number of ZZPhase gates in the Trotter circuit roughly scales $50+300(m-1)$ with the number of Trotter steps $m$. After the reduction, the simulation of 3 Trotter steps uses about 650 two-qubit gates.

After the reduction of two-qubit gate count, the memory noise from qubit idling becomes the dominant source of error, due to the large circuit depth. Although the $F$-moves commute with each other, the standard circuit compilation does not optimize circuit depth by rearranging the order of $F$-moves. To reduce the circuit depth, we arrange $F$-moves manually so that they can be applied in parallel.
Let us explain our strategy one by one. The $F$-moves in Eq.~\eqref{eq:action_trU} use external edges $c_1$, $c_2$, and $c_3$ as controlled qubits, and these edges are shared by two faces. Exploiting the freedom to rotate triangles by 120${}^\circ$, we apply $\prod_f\e^{-\im dt\, \tr U_{\tau}(f)}$ so that the number of control qubits used simultaneously by different plaquette implementations is minimized in each step of Eq.~\eqref{eq:action_trU}. Then, the $F$-moves that do not share controlled qubits can be applied in parallel.
Next, we explain the explicit gate ordering using the two plaquette terms as an example. We consider the Trotter evolution, $\e^{-\im dt \tr U_{\tau} (f_2)}\e^{-\im dt \tr U_{\tau} (f_1)}$. The naive circuit representation of $\e^{-\im dt \tr U_{\tau}(f_2)}\e^{-\im dt \tr U_{\tau}(f_1)}$ before the reordering can be written schematically as the left-hand side of 
\begin{equation}
\label{eq:Forder}
\begin{split}
    &\begin{quantikz}[column sep=.1cm, row sep=.2cm]
         & \gate[wires=5,disable auto height]{F}
         & 
         & 
         & 
         & \gate[wires=5,disable auto height]{F}
         &
         &
         & 
         & 
         & 
         &
         &
         \\
         &
         & 
         & 
         & 
         &
         &
         &
         &
         &
         &
         &
         &
         \\
         &
         & \gate[wires=4,disable auto height]{F_p}
         & \gate[wires=2,disable auto height]{\e^{-\im dt S}}
         & \gate[wires=4,disable auto height]{F_p}
         & 
         & 
         & 
         &
         &
         &
         & 
         & 
         \\
         &
         & 
         & 
         & 
         &
         &
         &
         & 
         & 
         & 
         &
         &
         \\
         &
         & 
         & 
         & 
         &
         &
         &
         & 
         & 
         & 
         &
         &
         \\
         &
         & 
         & 
         & 
         &
         &
         & \gate[wires=5,disable auto height]{F}
         & 
         & 
         & 
         & \gate[wires=5,disable auto height]{F}
         &
         \\
         &
         & 
         & 
         & 
         &
         &
         &
         &
         &
         &
         &
         &
         \\
         &
         & 
         & 
         & 
         &
         &
         &
         & \gate[wires=4,disable auto height]{F_p}
         & \gate[wires=2,disable auto height]{\e^{-\im dt S}}
         & \gate[wires=4,disable auto height]{F_p}
         & 
         & 
         \\
         &
         & 
         & 
         & 
         &
         &
         &
         & 
         & 
         & 
         &
         &
         \\
         &
         & 
         & 
         & 
         &
         &
         &
         & 
         & 
         & 
         &
         &
         \\
         &
         & 
         & 
         & 
         &
         &
         & 
         & 
         & 
         &
         &
         &
    \end{quantikz}
    \\
    &\hspace{2em}=    
    \begin{quantikz}[column sep=.1cm, row sep=.2cm]
         & \gate[wires=5,disable auto height]{F}
         & 
         & 
         & 
         & \gate[wires=5,disable auto height]{F}
         &
         \\
         &
         & 
         & 
         & 
         &
         &
         \\
         &
         & \gate[wires=4,disable auto height]{F_p}
         & \gate[wires=2,disable auto height]{\e^{-\im dt S}}
         & \gate[wires=4,disable auto height]{F_p}
         & 
         & 
         \\
         &
         & 
         & 
         & 
         &
         &
         \\
         &
         & 
         & 
         & 
         &
         &
         \\
         & \gate[wires=5,disable auto height]{F}
         &
         &
         &
         & \gate[wires=5,disable auto height]{F}
         & 
         \\
         &
         & 
         & 
         & 
         &
         &
         \\
         &
         & \gate[wires=4,disable auto height]{F_p}
         & \gate[wires=2,disable auto height]{\e^{-\im dt S}}
         & \gate[wires=4,disable auto height]{F_p}
         & 
         & 
         \\
         &
         & 
         & 
         & 
         &
         &
         \\
         &
         & 
         & 
         & 
         &
         &
         \\
         &
         & 
         & 
         & 
         &
         &
    \end{quantikz}    \;\;.
    \end{split}
\end{equation}
We reorder the gates to reduce the qubit-idling time as shown on the right-hand side of Eq.~\eqref{eq:Forder}. The reordering can be applied to all the plaquette terms so that the $F$-moves act in parallel. The effects of ordering can be quantified by the ZZPhase depth, that is, the number of ZZPhase gates in the longest acyclic path through the circuit from the qubit initializations to the final measurements. Using the reordering, the ZZPhase depth of the circuit is reduced from $140$ to $112$ for the two Trotter steps with $dt=1.0$, and from $271$ to $238$ for the three Trotter steps with $dt=1.0$. Notice that the depth may slightly vary for different $dt$ since we employ the variational circuit. 

Exploiting the aforementioned variational circuit reduction and gate reordering, we compute the regular Wilson loop at $t=0.5,\ldots,3.0$, and $3.75$. The Trotter step size $dt$ is carefully chosen to balance the Trotter error and noise effect because increasing the number of Trotter steps leads to less Trotter error at the cost of a larger influence of hardware noise. We simulate a single Trotter step with $dt=0.5$ and $1.0$ for computing the Wilson loop at $t=0.5$ and $1.0$. Similarly, we simulate two Trotter steps with $dt=0.75,1.0$ and $1.25$ for $t=1.5,2.0$ and $2.5$, and three Trotter steps with $dt=1.0$ and $1.25$ for $t=3.0$ and $3.75$. The executed circuits contain as many as 47 $F$-moves for simulating three Trotter steps. Notice that we can reuse qubits as ancilla qubits in the simulations of two Trotter steps, and we use $20$ qubits in total ($18$ qubits for the system and $2$ for ancilla qubits), while we use $26$ qubits in total ($18$ qubits for the system and $8$ for ancilla qubits) in the simulations of three Trotter steps. We take $200$ shots for each circuit.

\begin{figure}[t]
  \includegraphics[width=.49\textwidth]{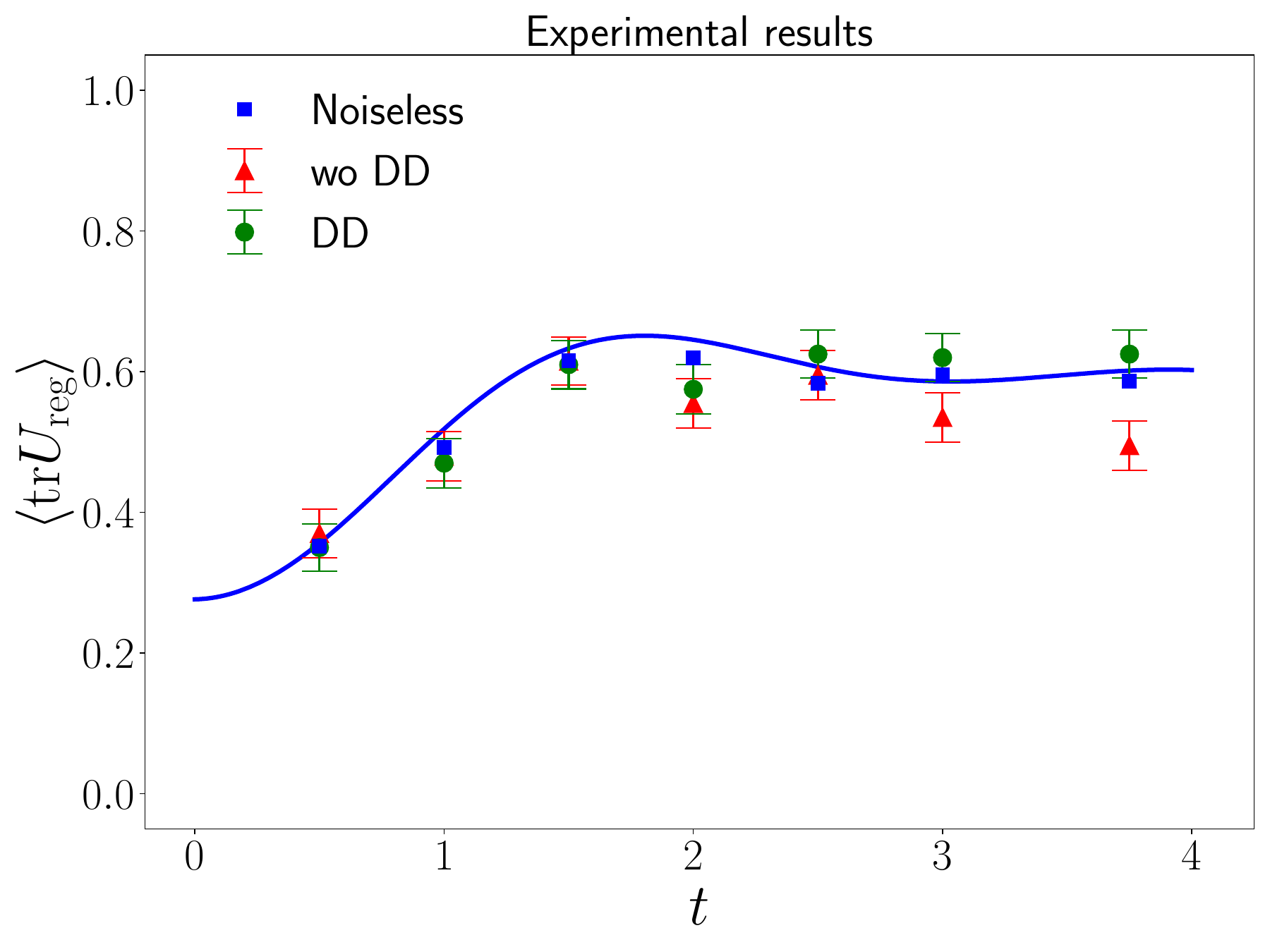}
  \caption{
    \label{fig:H2}
    Experimental results of the thermalization dynamics. Blue squares represent the classical noiseless simulations of the quantum circuits used in experiments. Red triangles and green circles represent experimental results without and with dynamical decoupling. The solid curves represent the classical noiseless simulations with a very small Trotter step, shown as the exact numerical simulations.
    }
\end{figure}%

The experimental results are shown in Fig.~\ref{fig:H2}. First, we see that the dynamics toward thermalization are successfully simulated using a real device. The raw data (red) starts to deviate from the noiseless values at $t=3.0$, which is simulated with 3 Trotter steps. However, the deviation is suppressed when DD is turned on, as shown by the green data points. This indicates that the dominant source of hardware noise is the coherent error that accumulates over time, which can be effectively mitigated by DD. Moreover, the agreement between the experimental data with DD and those of noiseless classical simulation within the statistical errors suggests that the incoherent gate error does not have a large impact on our circuits (notice that the two-qubit gate error of H2-1 is $1.05\times10^{-3}$, and then the fidelity of three Trotter steps is about $0.99895^{650}\sim0.5$ if we assume the depolarizing channel). This is attributed to the fact that our circuits are typically deep but acted upon by sparse gate operations, allowing coherent error to build up while qubits are idling. Such observations suggest that our circuits would benefit from more aggressive optimization based on their depth rather than two-qubit gate counts. Since such an optimization strategy has not been fully explored yet, we leave it as future work.

\section{Outlook}
\label{sec:conclusion}

We have demonstrated a digital quantum simulation of a nonabelian lattice gauge theory on a trapped-ion quantum computer. We construct a simplified model that preserves the nonabelian fusion rule of Wilson lines, and implement real-time evolution of the model with explicit $F$-moves describing the nonabelian anyons. Our experiments provide important lessons for scaling up quantum simulations of nonabelian lattice gauge theories based on $F$-moves toward regimes beyond the minimal truncation and the reach of classical computations.

The dominant error source in our circuits is attributed not to two-qubit gates but to memory noise accumulated during long multi-qubit controlled operations, that is, a large amount of Toffoli gates used in $F$-moves. This differs qualitatively from typical spin-model benchmarks, where error budgets track entangling-gate counts. We found that parallel scheduling of $F$-moves substantially reduces circuit depth, leading to significant accuracy improvements, combined with DD.

To further extend reachable times, we employ a variational circuit to compress the Trotter circuits. Instead of variationally approximating the whole Trotter circuit at once, we compress a sub-block in each Trotter step, consisting of one plaquette or one electric term. This modular approach not only reduces the classical optimization overhead but also allows reusing the same variational circuit for multiple blocks in larger systems.

Looking ahead, an important next step is extending our experiments to categories involving more complex anyons such as $\mathrm{SU}(2)_k$ or $\mathrm{SU}(3)_k$ with $k>3$. These theories are naturally represented by qudits and may require deeper circuits due to the increased complexity of their $F$-moves. In particular, $\mathrm{SU}(3)_k$ with $k>3$ exhibits fusion multiplicities, necessitating additional qubits to encode the Hilbert space of gauge fields~\cite{Hayata:2023bgh}. It is also important to consider theories with dynamical matter fields. Not only extending the complexity of the gauge group, but also scaling up the system size is crucial for accessing physically relevant regimes of nonabelian gauge theories. To this end, developing more efficient circuit decompositions of $F$-moves and Trotter circuits, together with error mitigation techniques tailored to nonabelian gauge theories, will be key to unlocking the full potential of quantum computers.

\section*{Data Availability}
 The experimental data that support the findings of this study are available at Zenodo~\cite{Zenodo}.

\section*{Code Availability}
 The code that supports the findings of this study is available from the corresponding author upon reasonable request.

\section*{Author contributions}
 T.H. and Y.H. conceived and designed the study. T.H. and Y.K carried out numerical studies and hardware experiments. T.H. and Y.K. analysed the data, and interpreted the results. T.H. wrote the manuscript with input from all authors.
 
\section*{Competing interests} 
The authors declare no competing financial or non-financial interests.
\\

\begin{acknowledgments}
We thank Kazuhiro Seki for collaboration on the early stages of this work. We also thank Andrew Potter, Enrico Rinaldi, and Yasuyoshi Yonezawa for their feedback on the manuscript. A portion of this work is based on results obtained from project JPNP20017, subsidized by the New Energy and Industrial Technology Development Organization (NEDO).  T.~H. is also supported by JSPS KAKENHI Grants No. JP24K00630 and No. JP25K01002. Y.~H. is supported by JSPS KAKENHI Grants No.~JP24H00975, No.~JP24K00630, No.~JP25K01002, and by JST, CREST Grant Number JPMJCR24I3. We are also grateful for the funding received from JST COI-NEXT (Grant No. JPMJPF2221).  
Additionally, we acknowledge the support from the UTokyo Quantum Initiative, and the RIKEN TRIP initiative (RIKEN Quantum).
\end{acknowledgments}

\onecolumngrid
\appendix

\section{Measurement of the Wilson loop operator}
\label{app:Wilson}
Notice that the $F$-moves are a basis transformation. We can measure the Wilson loop operator as we measure the Pauli $X$ by applying the Hadamard gate. Specifically, we measure $\tr U_\mathrm{P}$. The Wilson loop of other representations can be retrieved from the expectation value of $\tr U_\mathrm{P}$. For example, the regular Wilson loop is given as $\tr U_\mathrm{reg}=({1+\tr U_\mathrm{P}})/{2}$. We choose the basis in which the Wilson loop operator $\tr U_{P}$ acts on a tadpole graph:
\begin{equation}
\begin{split}
        \tr U_\mathrm{P}\vcenter{\hbox{\includegraphics[scale=0.25]{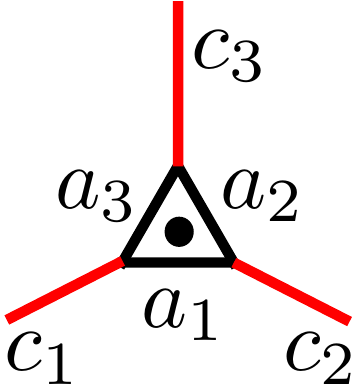}}} \;
                =
       \sum_{b_1,b_2,b_3}[F_{a_2}^{a_1c_1c_3}]_{a_3b_3}[F_{a_1}^{a_1b_3c_2}]_{a_2b_2}[S_\mathrm{P}^{b_2}]_{a_1b_1}
         \;\vcenter{\hbox{\includegraphics[scale=0.25]{out2_3.pdf}}}\;\;.
         \label{eq:basis_measure}
\end{split}
\end{equation}
Then, we measure $S_\mathrm{P}$ in its eigenbasis. Notice that $S^1_\mathrm{P}=R_y(\bar{\theta})Z Z R_y(\bar{\theta})Z$, and $S^\tau_\mathrm{P}=Z$. The measurement circuit is given as
\begin{equation}
       \parbox{4.cm}{
    \begin{quantikz}[column sep=.5cm, row sep=.2cm]
     \lstick{$|b_2\rangle$}
              & \gate{X}
         & \ctrl{1}
         &
\\
     \lstick{$|b_1\rangle$}    
              & \gate{Z}
         & \gate{R_y[\bar{\theta}]}
         & \meter{Z}
    \end{quantikz}       
       } \hspace{5em} ,
\end{equation}
where $|b_1\rangle$ and $|b_2\rangle$ qubits represent the qubits on the edges of the tadpole graph in Eq.~\eqref{eq:basis_measure}.

\section{Experimental details}
\label{app:experiment}

\subsection{Explicit construction of the first Trotter step}
\label{app:first_step}
We can explicitly compute the action of the Trotter gates to the initial state. Since we consider the 2nd order Trotter decomposition, the first Trotter gate applied to the initial state is $\prod_{f}\e^{\im \frac{\sqrt{5}Kdt}{4}\tr U_\mathrm{P}(f)}|\bm{0}\rangle$. It reads, for each face, 
\begin{equation}
\begin{split}
       \e^{\im \frac{\sqrt{5}Kdt}{4}\tr U_\mathrm{P}} \;
       \vcenter{\hbox{\includegraphics[scale=0.25]{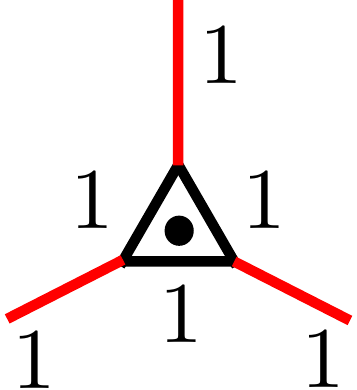}}}
      &=
       \sum_{a}\e^{\im \frac{\sqrt{5}Kdt}{4}[S_\mathrm{P}^1]_{a1}}
         \ \;\vcenter{\hbox{\includegraphics[scale=0.25]{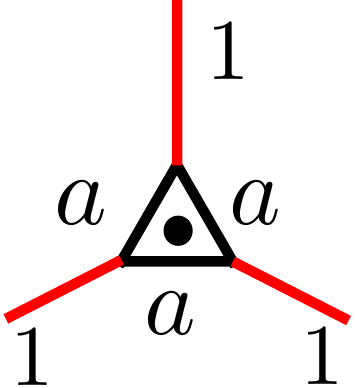}}}
         \;\;.
         \label{eq:action_trU_1st}
\end{split}
\end{equation}
This can be implemented as
\begin{equation}
       \vcenter{\hbox{
    \begin{quantikz}[column sep=.5cm, row sep=.2cm]
     \lstick{$|0\rangle$}    
              & \gate{Z}
         & \gate{R_y[\bar{\theta]}}
         & \gate{R_z[-\frac{\sqrt{5}Kdt}{2}]}
         & \gate{Z}
         & \gate{R_y[\bar{\theta]}}
         & \ctrl{1}
         & \ctrl{2}
         &
\\
     \lstick{$|0\rangle$}    
              & 
         & 
         & 
         & 
         & 
         & \gate{X}
         &
         &
         \\
     \lstick{$|0\rangle$}    
              & 
         & 
         & 
         & 
         & 
         & 
         & \gate{X}
         &
    \end{quantikz}       
       }} \;\; ,
\end{equation}
where qubits represent the anyons defined on the edges of the face. This circuit acts on the four faces in Eq.~\eqref{eq:faces}. We can completely remove the Toffoli gates associated with $F$-moves, which is crucial in the experiments using noisy quantum devices.

Next, we consider the evolution by the electric terms after Eq.~\eqref{eq:action_trU_1st}. We need to compute the diagrams such as
\begin{equation}
\begin{split}
       \e^{-\im \frac{dt}{2}E^2} \;
       \vcenter{\hbox{\includegraphics[scale=0.25]{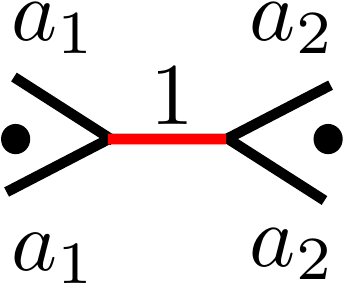}}} \;\;  .
         \label{eq:action_trE_1st}
\end{split}
\end{equation}
This diagram can also be computed explicitly as
\begin{align}
       \e^{-\im \frac{dt}{2}E^2} \;\;
       \vcenter{\hbox{\includegraphics[scale=0.28]{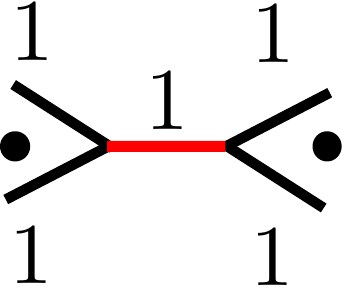}}}  
&\;= \hspace{3em}
       \vcenter{\hbox{\includegraphics[scale=0.28]{figures/1stE1.pdf}}} \;\;,
       \\
       \e^{-\im \frac{dt}{2}E^2} \;\;
       \vcenter{\hbox{\includegraphics[scale=0.28]{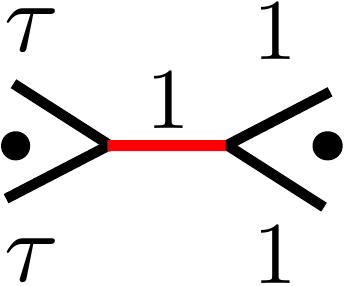}}}  
&\;= \hspace{1em}\e^{-\im \frac{dt}{2}} \;\;
       \vcenter{\hbox{\includegraphics[scale=0.28]{figures/1stE2.pdf}}} \;\;,
       \\
       \e^{-\im \frac{dt}{2}E^2} \;\;
       \vcenter{\hbox{\includegraphics[scale=0.28]{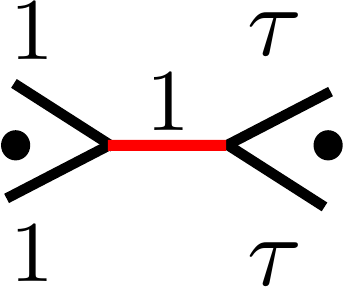}}}  
&\;= \hspace{1em}\e^{-\im \frac{dt}{2}} \;\;
       \vcenter{\hbox{\includegraphics[scale=0.28]{figures/1stE3.pdf}}} \;\;,
       \\
       \e^{-\im \frac{dt}{2}E^2} \;\;
       \vcenter{\hbox{\includegraphics[scale=0.28]{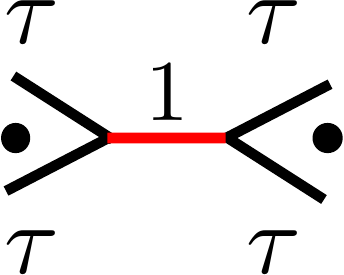}}}  
&\;= \sum_{f,b}[F_{\tau}^{\tau\tau\tau}]_{1f}[F_{\tau}^{\tau\tau\tau}]_{b f}
       \e^{-\im \frac{dt}{2} \delta_{f\tau}} \;\;
       \vcenter{\hbox{\includegraphics[scale=0.25]{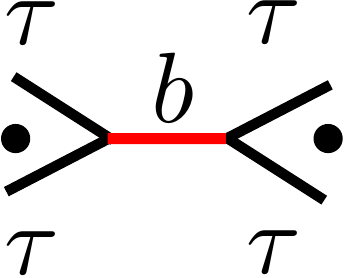}}}\;\; .
\end{align}
This can be implemented as
\begin{equation}
       \parbox{8.cm}{
    \begin{quantikz}[column sep=.5cm, row sep=.2cm]
     \lstick{$|a_1\rangle$}
              & \ctrl{2}
         & 
         & 
         & 
         & \ctrl{2}
         & 
         & 
         & \ctrl{1}
         & 
         &
         &
\\
     \lstick{$|a_2\rangle$}    
              & 
         & \ctrl{1}
         & 
         & \ctrl{1}
         & 
         & 
         & 
         & \ctrl{1}
         & 
         &
         &
         \\
     \lstick{$|0\rangle$}    
              & \gate{X}
         & \gate{X}
         & \gate{P[-\frac{dt}{2}]}
         & \gate{X}
         & \gate{X}
         & \gate{R_y[2\theta]} 
         & \gate{X}
         & \gate{P[-\frac{dt}{2}]}
         & \gate{X}
         & \gate{R_y[-2\theta]} 
         &
    \end{quantikz}       
       } \hspace{20em} ,
\end{equation}
where $|a_1\rangle$ and $|a_2\rangle$ qubits represent two of four legs in Eq.~\eqref{eq:action_trE_1st}. This circuit acts on the six edges in Eq.~\eqref{eq:edges2}. The naive implementation based on Eq.~\eqref{eq:action_E2_exp} consumes twelve Toffoli gates, but we can reduce them to only one controlled-controlled phase gate. This is also crucial on the experiments using noisy quantum devices.

\subsection{Explicit construction of the general Trotter steps}
\label{app:CPFlow}
We can write the time evolution by magnetic terms as
\begin{equation}
\begin{split}
\e^{\im \frac{\sqrt{5}Kdt}{2}\tr U_\mathrm{P}}
     \vcenter{\hbox{\includegraphics[scale=0.25]{figures/in2.pdf}}}
      &=
       \sum_{b_1,b_2,b_3,b_2^\prime,b_3^\prime}[F_{a_2}^{a_1c_1c_3}]_{a_3b_3}[F_{a_1}^{a_1b_3c_2}]_{a_2b_2}[\e^{\im \frac{\sqrt{5}Kdt}{2} S^{b_2}_\mathrm{P}}]_{a_1b_1}
       [F_{b_1}^{b_1b_3c_2}]_{b_2^\prime b_2}[F_{b_2^\prime}^{b_1c_1c_3}]_{b_3^\prime b_3}
         \ \;\vcenter{\hbox{\includegraphics[scale=0.25]{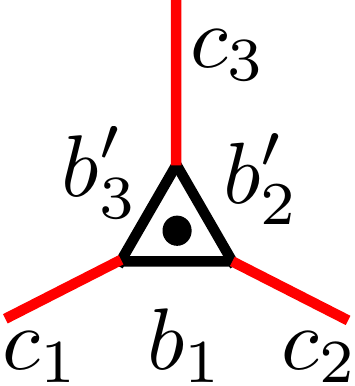}}}
        \;\;.
         \label{eq:action_general_B}
\end{split}
\end{equation}
Since $a_1=a_2=a_3=a$ after the first Trotter step, we can replace the general $F$-move $[F_{a_2}^{a_1c_1c_3}]_{a_3b_3}$ with the simpler one for the magnetic evolution in the second Trotter step. However, we can no longer reduce the circuit by using information of the initial condition for subsequent Trotter gates.

For further reduction, we replace the original circuit with the variational (much shallower) one. To this end, we employ CPflow~\cite{Nemkov:2022mah}. The time evolution operator generally has the form $U=F\cdots F \e^{i dt \Lambda} (F\cdots F)^\dagger$ with $\Lambda$ being $S$ for the magnetic evolution, and $Z$ for the electric evolution.  Notice that the basis transformation and re-transformation by $F$-moves, $F\cdots F$ and $(F\cdots F)^\dagger$ consume many two-qubit gates, and also result in the deep circuit due to the successive controlled operation. However, $U$ may be almost trivial unitary if $dt$ is sufficiently small (the most parts may be the trivial basis transformation and re-transformation). This fact leads us to variationally approximate the time evolution operator with a much shallower circuit, keeping the accuracy. We keep the loss of the CPflow, so that the difference between the original and variational circuits is negligibly small compared with other errors such as the Trotter error and device noise. Specifically, we set loss in the CPflow to $5\times10^{-4}$. In our case, the CPflow finds the variational circuit efficiently if the number of involved qubits is less than $6$. Therefore, we replace part of the magnetic evolution operator with the variational circuit:
\begin{equation}
\begin{split}
       [F_{a_1}^{a_1b_3c_2}]_{a_2b_2}[\e^{\im \frac{\sqrt{5}Kdt}{2}\tr S^{b_2}_\mathrm{P}}]_{a_1b_1}
       [F_{b_1}^{b_1b_3c_2}]_{b_2^\prime b_2}
       \simeq [U_\mathrm{CP}]
\end{split}
\end{equation}
and the magnetic evolution operator is schematically written as
\begin{equation}
\begin{split}
\e^{\im \frac{\sqrt{5}Kdt}{2}\tr U_\mathrm{P}}
     \vcenter{\hbox{\includegraphics[scale=0.25]{figures/in2.pdf}}}
      &=
       \sum_{b_1,b_2,b_3,b_2^\prime,b_3^\prime}[F_{a_2}^{a_1c_1c_3}]_{a_3b_3}[U_\mathrm{CP}][F_{b_2^\prime}^{b_1c_1c_3}]_{b_3^\prime b_3}
         \ \;\vcenter{\hbox{\includegraphics[scale=0.25]{figures/out2_5.pdf}}}
         \;\;.
         \label{eq:action_general_B_CP}
\end{split}
\end{equation}
Notice that we apply the first and last full $F$-moves without approximation using ancilla qubits. On the other hand, the whole electric evolution operator can be approximated by the variational circuit at once.


\bibliography{ym}
\end{document}